\documentclass[fleqn,usenatbib]{mnras}

\usepackage[T1]{fontenc}
\usepackage{ae,aecompl}

\usepackage{graphicx}	
\usepackage{amsmath}	
\usepackage{amssymb}	
\usepackage{ulem}
\usepackage{epsfig} 

\title[Spitzer GW170817]{Spitzer Mid-Infrared Detections of Neutron Star Merger GW170817 
Suggests Synthesis of the Heaviest Elements}

\author[M. M. Kasliwal et al.]{Mansi M. Kasliwal,$^{1}$\thanks{E-mail: mansi@astro.caltech.edu}
Daniel Kasen$^{2,3}$, 
Ryan M. Lau$^{1}$, 
Daniel A. Perley$^{4}$,  
\newauthor
Stephan Rosswog$^{5}$, 
Eran O. Ofek$^{6}$, 
Kenta Hotokezaka$^{7}$, 
Ranga-Ram Chary$^{8}$, 
\newauthor
Jesper Sollerman$^{9}$, 
Ariel Goobar$^{10}$ 
and David L. Kaplan$^{11}$
\\
$^{1}${Division of Physics, Mathematics and Astronomy, California Institute of Technology, Pasadena, CA 91125, USA}  \\
$^{2}${Department of Astronomy and Department of Physics, University of California, Berkeley, CA 94720-3411, USA } \\
$^{3}${Lawrence Berkeley National Laboratory, 1 Cyclotron Road, MS 50B-4206, Berkeley, CA 94720, USA} \\
$^{4}${Astrophysics Research Institute, Liverpool John Moores University, IC2, Liverpool Science Park, 146 Brownlow Hill, Liverpool L3 5RF, UK} \\
$^{5}${The Oskar Klein Centre,  Department of Astronomy, Stockholm University, AlbaNova, SE-106 91 Stockholm, Sweden} \\
$^{6}${Department of Particle Physics \& Astrophysics, Weizmann Institute of Science, Rehovot 7610001, Israel} \\
$^{7}${Department of Astrophysical Sciences, 4 Ivy Lane Princeton University, Princeton, NJ 08544} \\
$^{8}${Infrared Processing and Analysis Center, Caltech, Pasadena CA} \\
$^{9}${The Oskar Klein Centre,  Department of Astronomy, Stockholm University, AlbaNova, SE-106 91 Stockholm, Sweden} \\
$^{10}${The Oskar Klein Centre, Department of Physics, Stockholm University, AlbaNova, SE-106 91 Stockholm, Sweden} \\
$^{11}${Department of Physics, University of Wisconsin, Milwaukee, WI 53201, USA}
}

\date{Accepted XXX. Received YYY; in original form ZZZ}

\pubyear{2018}

\begin{document}
\label{firstpage}
\pagerange{\pageref{firstpage}--\pageref{lastpage}}
\maketitle

\begin{abstract}
We report our Spitzer Space Telescope observations and detections of the binary neutron star merger GW170817. At 4.5$\mu$m, GW170817 is detected at 21.9\,mag AB at +43\,days and 23.9\,mag AB at +74\,days after merger. At 3.6$\mu$m, GW170817 is not detected to a limit of 23.2\,mag AB at +43\,days and 23.1\,mag AB at +74\,days. 
Our detections constitute the latest and reddest constraints on the kilonova/macronova emission and composition of heavy elements. The 4.5$\mu$m luminosity at this late phase cannot be explained by elements exclusively from the first abundance peak of the r-process. Moreover, the steep decline in the Spitzer band, with a power-law index of 3.4 $\pm$ 0.2,
can be explained by a few of the heaviest isotopes in the third abundance peak with half-life around 14\,days dominating the luminosity  (e.g.  $^{140}$Ba, $^{143}$Pr, $^{147}$Nd, $^{156}$Eu, $^{191}$Os, $^{223}$Ra, $^{225}$Ra, $^{233}$Pa, $^{234}$Th) or a model with lower deposition efficiency. This data offers evidence that the heaviest elements in the second and third r-process abundance peak were indeed synthesized. Our conclusion is verified by both analytics and network simulations and robust despite intricacies and uncertainties in the nuclear physics. Future observations with Spitzer and James Webb Space Telescope
will further illuminate the relative abundance of the synthesized heavy elements. 
\end{abstract}

\begin{keywords}
stars: neutron, stars: black holes, gravitational waves, nucleosynthesis
\end{keywords}



\section{Introduction}

The discovery of gravitational waves from merging neutron stars, GW170817 \citep{GW170817}, 
offered the first opportunity to directly test the long-standing hypothesis of whether 
these are the sites of heavy element production \citep{Lattimer1974}.  The discovery of 
long-lived infrared emission from GW170817 has provided unequivocal evidence that 
these are indeed prolific sites of r-process nucleosynthesis \citep{Coulter17,Drout17,Evans17,Kasliwal17c,Smartt17,SoaresSantos2017,Cowperthwaite17,Arcavi2017}.
The rapid photometric evolution to the redder wave-bands and the sustained luminous infrared emission for a few weeks was consistent with predictions from a suite of kilonova/macronova models \citep{Li1998,Kulkarni2005,Metzger2010,Barnes2013,Kasen2013,Tanaka2013, Wollaeger2017,Rosswog2017}. The photospheric infrared spectra showed possible evidence of lanthanides such as Neodymium \citep{Kasen17}.

While there is an emerging consensus in the literature that $\approx$0.04--0.05 M$_{\odot}$ of heavy elements was synthesized and moving at $\approx$0.1--0.3c, 
there is much debate on which of the heavy elements were synthesized and whether the abundance distribution matches solar. The solar heavy element distribution has three distinct abundance peaks between atomic mass numbers 70 and 200: the first abundance peak spans 70--88, the second  peak spans 120--140, lanthanides span 139--180 and the third  peak spans 180--200. The photometric data has been explained both with and without the third r-process peak including the heaviest elements \citep{Rosswog2018, Smartt17, Waxman17}. The very red spectral energy distribution suggests a non-zero lanthanide fraction, which in-turn would suggest that elements at all three r-process peaks are synthesized \citep{Kasen17,Pian2017}. 

Despite an intensive campaign by a suite of telescopes worldwide, the infrared monitoring of GW\,170817 came to a grinding halt three weeks 
post-merger due to the target becoming too close to the sun. Here, we report late-time infrared photometry from the warm {\it Spitzer} Space
Telescope \citep{Werner2004}, the only telescope that was able to collect infrared data at +43\,d and +74\,d after merger despite the proximity to the sun. We use {\it Spitzer} observations at +264\,d as a reference for image subtraction analysis. Our photometry is inconsistent with that reported in \citealt{Villar18}. 

\section{Observations}

\begin{figure*} 
\centering
\includegraphics[width=0.7\textwidth]{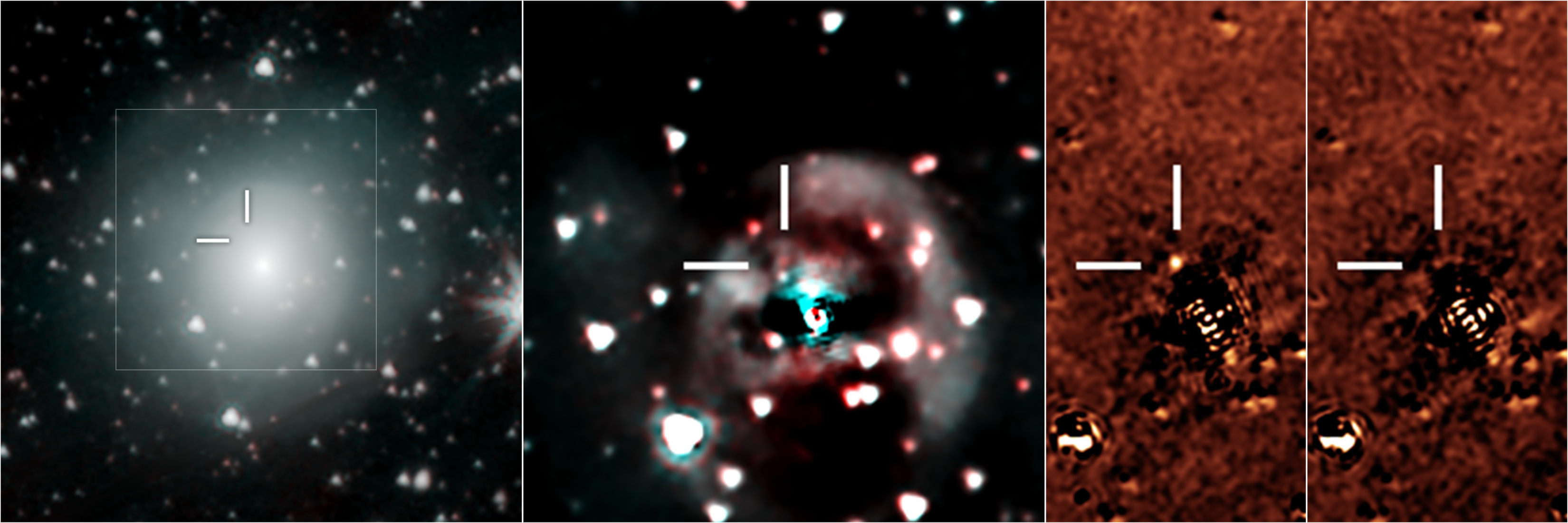}
\caption{{\it Panel 1:} Combined Spitzer 3.6$\mu$m and 4.5$\mu$m image, depicting that the faint transient GW\,170817 is buried in the bright host galaxy NGC\,4993. {\it Panel 2:} Subtracting the galaxy light by fitting a GALFIT model clearly shows the red transient in the first epoch image, +43d after merger.  
{\it Panel 3a:} Proper image subtraction of Epoch 3 reference data from Epoch 1 using the ZOGY algorithm boosts the S/N of our detection of GW170817. {\it Panel 3b:} ZOGY subtraction of Epoch 3 reference data from Epoch 2 yields a marginal detection of GW170817. The orientation of all four panels is such that North is up and East is left. 
The dimensions of the panels are 2.75$\arcmin$ $\times$ 2.75$\arcmin$, 1.38$\arcmin$ $\times$ 1.38$\arcmin$, 0.69$\arcmin$ $\times$ 1.38$\arcmin$ and 0.69$\arcmin$ $\times$ 1.38$\arcmin$.
\label{fig:sub}}
\end{figure*}

We observed GW170817 thrice with the InfraRed Array Camera (IRAC; \citealt{Fazio2004}) aboard the warm {\it Spitzer} Space Telescope at the beginning and end of the first visibility window after explosion and again in the second visibility window (PID 13202, PI Kasliwal). Each epoch constituted a 10\,hr integration split into 30\,s to minimize 
galaxy core saturation. Due to the larger data volume, each epoch was split into two back-to-back observing blocks. Each epoch had observations in both the 3.6$\mu$m filter and 4.5$\mu$m filter. The first epoch on 2017 Sep 29 was +43\,d after merger, the second epoch on 2017 Oct 30 was +74\,d after merger and the third epoch on 2018 May 8 was +264\,d after merger. Archival imaging of NGC\,4993 also exists from 2014 (PID 10043, PI K. Sheth, we stacked 2014-10-12 and 2014-09-12 data). Data was reduced and mosaiced by the IRAC pipeline. 

The IRAC point-spread function is complex: although its core is compact, significant light is scattered far from the position of an object in a complex, asymmetric pattern. The fixed detector/optical diffraction patterns in the IRAC PSF profiles complicates identification of sources in the vicinity of the bright galaxy core since they cannot be easily matched and subtracted between observations taken at different position angles. Specifically, the position angles were 106\,deg for Epoch 1, 114\,deg for Epoch 2, $-$65\,deg for Epoch 3 and 108\,deg for the archival observation. Given the complexity of the underlying galaxy background and the mismatched position angles, we undertook several independent data analysis methods that facilitated multiple consistency checks. 

We first describe our preferred method that yielded the highest Signal to Noise (S/N) detections presented in Table~\ref{tab:data}.
To remove the flux from the bright galaxy, we employ the software
package GALFIT \citep{Peng2010}, interfaced via a custom python
wrapper for galaxy subtraction \citep{Perley2016}.  We use the
post-Basic Calibrated Data (PBCD) images from the Spitzer archive for each set of observations (split into two
observing blocks for each Epoch).  We used the first reference observation to fit a
Sersic model for the host galaxy profile (simultaneously fitting the 20 brightest stars within the fitting box).  The Point Response Function (PRF) files
for both Spitzer/IRAC bands were downloaded from the Spitzer website and
used as the PSFs, and our fit is restricted to a 58x58 arcsec box around
the galaxy centroid.  A reasonable fit is obtained for this model and
most of the galaxy light and foreground starlight is effectively
removed (Figure~\ref{fig:sub}). We see some asymmetric residuals which visually
match the structure of the tidal 'shells' visible in HST imaging of the
field as well as some residuals from much fainter foreground stars and
background galaxies.  At 4.5$\mu$m, We measure a
Sersic index of 4.24, a half-light radius of 10.74 arcseconds, an
integrated magnitude of 12.29 (AB), and an axis ratio of 0.849. At 3.6$\mu$m, we measure a Sersic index of 4.65, a half-light radius of 12.38 arcseconds, 
axis ratio of 0.838 and integrated magnitude of 11.65 AB. 
We then repeat this procedure for all other PBCD images, but freeze the
fundamental galaxy fit parameters to their values above to ensure a
consistent subtraction across every image.  We allow the PA and
centroid location to vary to allow for astrometric inaccuracies.   This
effectively cleans the galaxy light and its complex scattering pattern
consistently for all images, with the exception of the faint tidal rings. 

To remove this residual light, and other variations not well-fit by the Sersic model, we used the proper image subtraction routine for optimal transient detection and photometry described in \citealt{zogy} (ZOGY). We used a bright nearby star, 2MASS 13094158-2323149 \citep{2MASS}, as our reference for the ZOGY PSF and measure the flux-correction factors from the ZOGY output of the difference-image PSF. Following the guidelines in the IRAC Handbook, our correction factors are normalized to a 20\,pix radius which includes 100\% of the flux (see Table~2). As a consistency check, we compute three apertures (radii of 4, 6 and 10\,pix) that match those in the IRAC Handbook and get similar correction factors. 

We applied ZOGY to subtract Epoch 3 images from each of the Epoch 1 and Epoch 2 images (Figure~\ref{fig:sub}). The advantage of using Epoch 3 as a reference for image subtraction over archival data is that it is both comparable in depth to Epoch 1/Epoch 2 and devoid of transient light. As an additional consistency checks, we also applied ZOGY to subtract archival Spitzer data of NGC\,4993 taken in 2014 from the new data as the archival images are better matched in position angle than the Epoch 3 reference. Additionally, we applied ZOGY to subtract Epoch 2 from Epoch 1 as a consistency check on the difference in flux.  

All methods reveal a source at 4.5$\mu$m in Epoch 1 and Epoch 2. All methods reveal no source at 3.6$\mu$m in either Epoch 1 or Epoch 2. Each of these difference images were subject to both PSF-fit and aperture photometry tasks. 
We summarize the photometry and 3$\sigma$ upper limits measured from our ZOGY subtractions via PSF-fit photometry in Table~\ref{tab:data}.  (Paranthetical errors added to the measured magnitudes are on account of our analysis of the noisy residuals from stars in the subtracted image suggesting that there may be an additional systematic error of 0.05 mag). We note that our Epoch 1 photometry is inconsistent and brighter by 1 magnitude compared to that reported in \citealt{Villar18}. We undertake the following consistency checks. 
\begin{itemize}
\item We get consistent magnitudes for aperture photometry and PSF photometry. For a 2.5\,pix aperture, the aperture magnitudes are 21.99 $\pm$ 0.04 for Epoch 1 and 24.14 $\pm$ 0.30 AB mag for Epoch 2, consistent with the results from PSF-fit photometry in Table~\ref{tab:data}. 

\item The sum of PSF-fit fluxes of the (Epoch 1 - Epoch 2) and (Epoch 2 - Epoch 3) difference images equal the the flux in the (Epoch 1 - Epoch 3) difference image. Specifically, the sum of the measured flux in the first two difference images is (5.47 $\pm$ 0.14 $\mu$Jy) + (1.04 $\pm$ 0.21 $\mu$Jy), which is consistent with the measured flux in the (Epoch 1 - Epoch 3) difference image of (6.39 $\pm$ 0.21 $\mu$Jy).

\item If we increase the Gaussian FWHM of the PSF-fit to 3.5\,pix and apply the appropriate correction factor, we measure a magnitude of 21.93 $\pm$ 0.06, consistent with the 2.8\,pix FWHM measurement at Epoch 1. 

\item We get consistent fluxes for Epoch 1 using either the shallow archival reference or the deeper Epoch 3 reference. The PSF-magnitude of Epoch 1 in the archival difference is 21.79 $\pm$ 0.09 AB mag, consistent with the late-time difference albeit with larger error bars.  

\item We get consistent fluxes for Epoch 1 if we directly apply ZOGY to subtract Epoch 3 without first applying the GALFIT-model. We derive 21.94 $\pm$ 0.25\,mag. The subtraction is noisier by direct subtraction, hence, we prefer the two-step method described above. 

\item We re-do aperture corrections with a different sky annulus (5--7\,pix) and scaling the ZOGY PSF to the standard PRF after re-normalizing the sky. We also take into account color corrections for this red source by multiplying the measured 4.5$\mu$\,m flux by 1.024 (and 3.6$\mu$\,m flux by 1.0614) . This gives 21.92 $\pm$ 0.09\,mag at Epoch 1 and 23.94 $\pm$ 0.4\,mag at Epoch 2, consistent with Table~\ref{tab:data}.

\end{itemize}

Converting to flux density, we get F$_{\nu}$ = 6.43 $\times$ 10$^{-29}$ erg s$^{-1}$ cm$^{-2}$ Hz$^{-1}$ at Epoch 1 and F$_{\nu}$ = 1.04$\times$ 10$^{-29}$ erg s$^{-1}$ cm$^{-2}$ Hz$^{-1}$ at Epoch 2. Now, $\Delta{\nu}$-L$_{\nu}$ would be a strict lower limit on the total bolometric luminosity. If we assume a power-law ${\nu}$-L$_{\nu}$ approximation to bolometric, the assumed correction factor is the ratio between the central frequency and bandwidth i.e. a multiplicative factor of 4.3 (since Channel 2 of {\it Spitzer/IRAC} spans 3.955$\mu$m to 5.015$\mu$m). 

At this late phase, we expect optically thin, nebular conditions and a blackbody approximation with a photosphere is unlikely to be applicable. Nevertheless, we proceed with blackbody calculations as another way to estimate the bolometric correction. The observed Spitzer/IRAC color ([4.5] - [3.6]) of  1.3\,mag suggests a blackbody temperature of 420\,K at Epoch 1 (the Epoch 2 color is not constraining). This suggests a multiplicative bolometric correction factor of $\approx$16. In the rest of the paper, we assume a ${\nu}$-L$_{\nu}$ approximation to the bolometric luminosity of 7.8$\times$10$^{38}$ erg s$^{-1}$ at Epoch 1 and 1.3$\times$10$^{38}$ erg s$^{-1}$ at Epoch 2.  

We check whether synchrotron emission could contribute to the observed flux. Assuming the spectral index presented in \citet{Mooley18}, and a flux density of 44$\mu$Jy at 3\,GHz measured at the same phase,  we estimate that the synchrotron contribution at 4.5$\mu$m would be 1.1$\times$10$^{-30}$ erg s$^{-1}$ cm$^{-2}$ Hz$^{-1}$ at Epoch 1. This is $\approx$60 times smaller than the observed flux density and hence, we conclude that the synchrotron contribution is negligible.

\section{Implications on abundances of r-process elements}

At the epochs of the Spitzer observations ($t \gtrsim 40$~days) the ejecta of kilonovae are
expected to be optically thin to optical/infrared photons. 
The bolometric luminosity should then be independent of viewing angle and 
follow the instantaneous radioactive heating rate,
 $L(t) \approx M_{\rm ej} \dot{\epsilon}(t) f(t)$ where $M_{\rm ej}$ is
 the ejecta mass, $\dot{\epsilon}(t)$ the radioactive power per gram, and
 $f(t)$ the efficiency with which radioactive energy is thermalized.
The late-time Spitzer data can thus be used to derive constraints on the ejecta mass and composition 
that are independent of the complex ejecta  opacity and geometry. The main limitation is
the uncertain bolometric corrections.
 
 The radioactive power of r-process matter is often described by
 a power-law, $\dot{\epsilon}(t) \propto t^{-4/3}$, which is the behavior of a statistical distribution of isotopes with
beta-decay half-lives roughly equally distributed in log time. 
The thermalization efficiency for such an isotopic distribution is approximately \citep{Kasen18}
\begin{equation}
f(t)  \approx p_\gamma (1 - e^{-t_\gamma^2/t^2}) + p_e (1 + t/t_e)^{-n},
\end{equation}
where $p_\gamma \approx 0.4, p_e \approx 0.2$ are the fraction of beta-decay energy emitted as  gamma-rays and electrons, respectively.
For ejecta masses and velocities in the range $M \approx 0.01 - 0.05~M_\odot, v \approx 0.1c-0.2c$ the timescale for gamma-rays to become inefficient to thermalization is $t_\gamma \approx 0.5 - 2$~days while  that for electrons is $t_e \approx 10 - 40$~days.  The exponent $n \approx 1$ for typical conditions, though $n$ can be larger depending on the details
of the thermalization and decay physics \citep{Kasen18}. 

Figure~\ref{fig:heating} shows calculations of the radioactive power $\dot{\epsilon}(t)$  derived from
 detailed r-process nuclear reaction
networks for outflows with a range of physical conditions (initial electron fractions $Y_e = 0.05 - 0.5$, expansion velocity of 0.2c, ejecta mass of 0.05 M$_{\odot}$ \citealt{Rosswog2018}).  
At +43\,d, the radioactive power ranges from 
$\dot{\epsilon} \approx 0.5 - 2.5 \times 10^{8}~{\rm erg~s^{-1}~g^{-1}}$.
Adopting the  $\nu L_\nu$  luminosity at epoch 1 of $L_{43} = 7.8 \times 10^{38}~{\rm erg~s^{-1}}$
and using an efficiency factor $f = 0.1$ (appropriate for  $t_e \approx 30$~days) implies an ejecta mass of 
$M_{\rm ej} \approx 1.6 - 7.8  \times 10^{-2}~M_\odot$. Within large uncertainties, the mass range is consistent with
that inferred from analysis of early time observations of GW170817 
\citep{Coulter17,Drout17,Evans17,Kasliwal17c,Smartt17,SoaresSantos2017,Cowperthwaite17,Arcavi2017}, and
provides additional evidence that the neutron star merger produced a large quantity of radioactive ejecta.
 
Between the two epochs of Spitzer observations, the luminosity dropped by a factor $L_1/L_2 \approx 6.2$
corresponding to a power-law L $\propto$ t$^{-3.4 \pm 0.2}$.
This is steeper than the $L  \propto t^{-7/3}$ dependence of
statistical distribution of isotopes with power $\dot{\epsilon} \propto t^{-4/3}$ with inefficient thermalization  
$f(t) \propto t^{-1}$. Alternately, the observed decline can be explained if the efficiency drops even more rapidly, $f(t) \propto t^{-2}$, as suggested by \cite{Waxman17} 
(although such a steep dependence of $f(t)$ is not consistent with the numerical thermalization calculations of \citep{Barnes2016}). Based on late-time optical 
data, \citealt{Waxman17} and  \citealt{Arcavi2018} also suggested a similarly steep late-time power-law slope of  t$^{-3}$. 

It is possible that the decline in luminosity between the two Spitzer epochs is a result of the 
spectral energy progressively moving out of 4.5$\mu$m band, such that the bolometric correction increases with time.
If such a color evolution occurred, the spectrum must have moved redward of $5~\mu$m, 
as the upper limits in the 3.6$\mu$m band rule out a substantial increase of the flux at bluer wavelengths. 

If we assume, on the other hand, that the bolometric correction remained largely unchanged between the two epochs, the two Spitzer epochs suggest that the underlying radioactivity has 
deviated from the $\dot{\epsilon} \propto t^{-4/3}$ power-law behavior.  This is expected to occur when the decay becomes dominated by one or a few isotopes rather than a statistical distribution \citep{Kasen18,Wu2018}. For a single dominant isotope the energy generation rate  follows
$\dot{\epsilon}(t)  \propto e^{-t/t_i}$ where $t_i$ is the decay timescale. 
Taking into account the effects of inefficient thermalization, the heating from a single isotope at times
 $t \gtrsim t_e$ is \citep{Kasen18}
\begin{equation}
L \propto 
\frac{\exp \left[ - 
\sqrt[3]{ 3 t/2 t_e} (t_e/t_i)  \right]}{(t/t_e)^{7/3}}.
\label{eq:q_ia}
\end{equation}
From Equation~\ref{eq:q_ia} and using $t_e = 30$~days the observed ratio $L_1/L_2 \approx 6.2$ implies
heating dominated by an isotope with decay time $t_i \approx 14$~days.

\begin{figure} 
\centering
\includegraphics[width=0.5\textwidth]{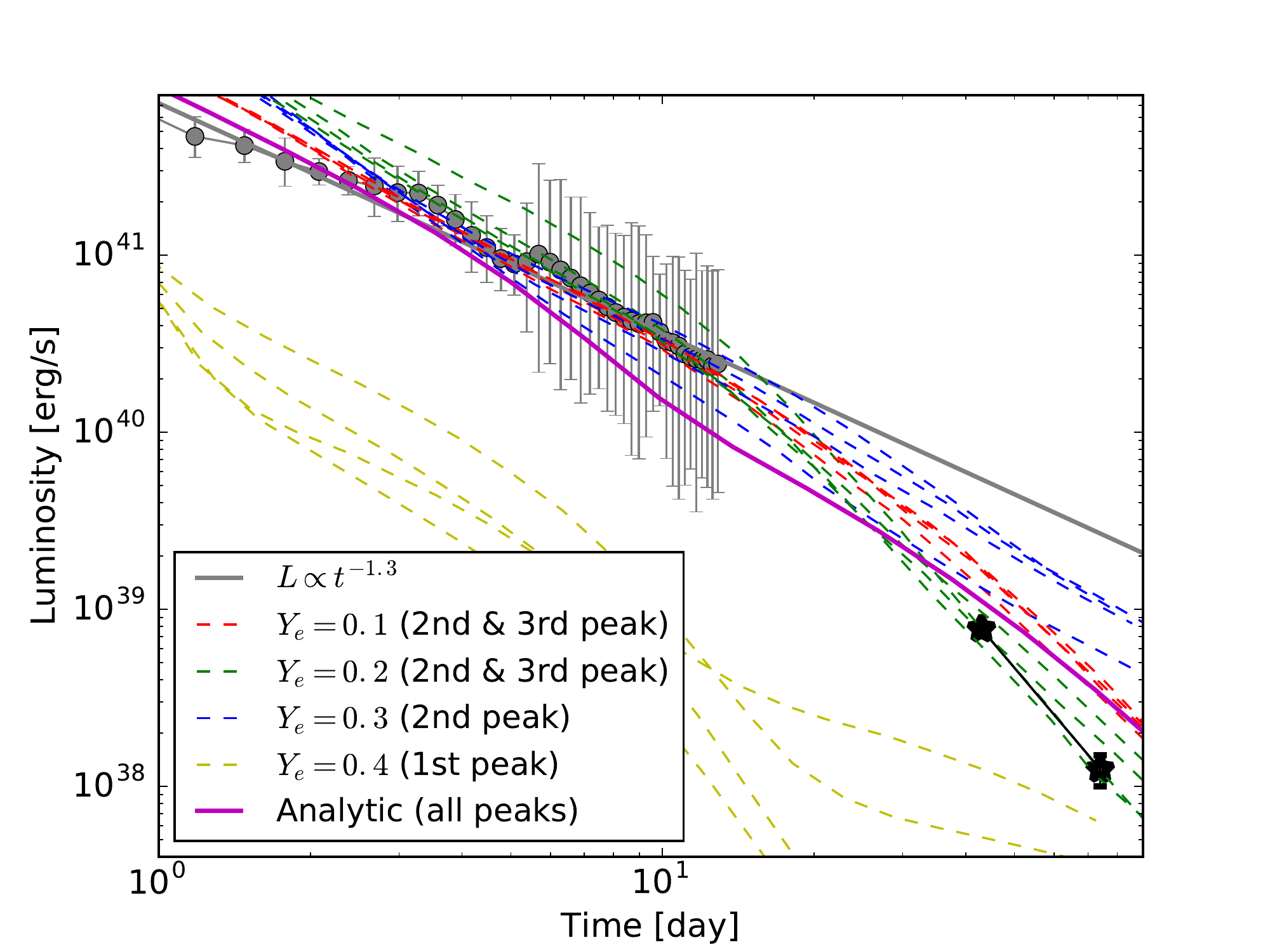}
\caption{Comparing early-time bolometric data (circles, \citealt{Kasliwal17c}) and late-time Spitzer detections (stars, this paper) with the predicted radioactive luminosity as a function of time (lines). The dashed colored lines show a luminosity L = M$_{\rm ej}$ $\dot{\epsilon}(t)$ f(t), where the ejecta mass M$_{\rm ej}$ = 0.05 M$_{\odot}$, the thermalization efficiency f(t) is from \citealt{Kasen18}, and the radioactive power $\dot{\epsilon}(t)$ is from the detailed nuclear reaction network calculations of \citealt{Rosswog2018}. $\dot{\epsilon}(t)$ explores a range of electron fraction Ye and expansion velocity from 0.1c to 0.4c. Outflows with Ye$<$0.25 synthesize the heaviest r-process elements in the second-peak and third-peak and show a steeper late time decline, whereas those with Ye$\gtrsim$0.25 produce relatively lighter elements and have a shallower decline due to the presence of longer lived radioactive isotopes. Also shown is the power law inferred from early-time data (gray solid line) and an analytic estimate of beta decay rates assuming a statistical distribution (magenta solid line; \citealt{Hotokezaka17}). 
\label{fig:kasen}}
\end{figure}

\begin{figure} 
\centering
\includegraphics[width=0.5\textwidth]{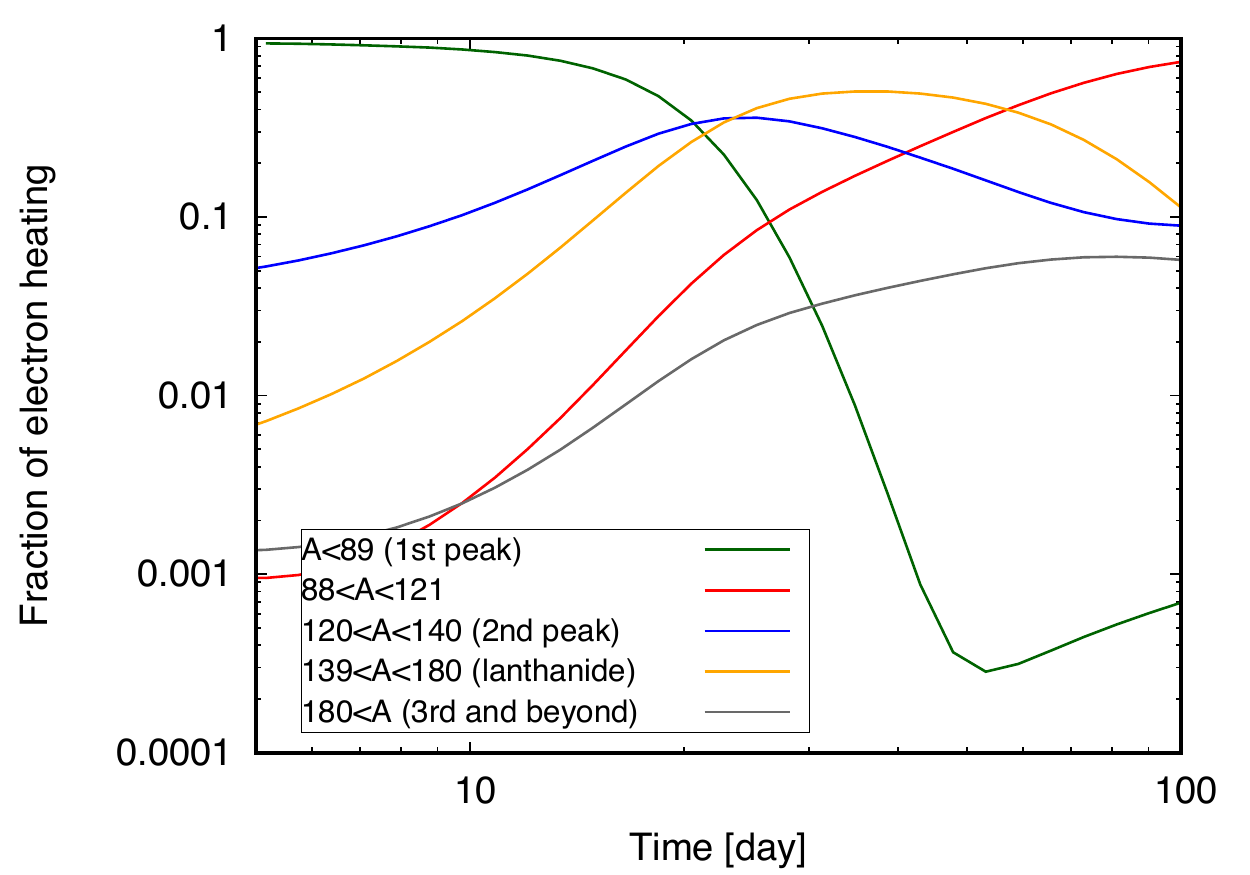}
\caption{ 
Fraction of electron heating contributed by various sets of elements as a function of time using the solar abundance pattern including the first peak. While the first peak dominates at early-time, our detection of late-time emission requires elements in the second and third peak. 
\label{fig:fracheating}}
\end{figure}

If the late time radioactivity is indeed dominated by a single isotope, this provides constraints on the ejecta composition.  For merger outflows with electron fractions $Y_e \lesssim 0.25$ the nucleosynthesis proceeds to the 3rd r-process peak (Figure~\ref{fig:heating}) and the radioactive power $\dot{\epsilon}(t)$  steepens at times $t \gtrsim 40$~days to a decline rate consistent with the two Spitzer epochs (Figure~\ref{fig:kasen}).  For electron fractions $Y_e \gtrsim 0.25$, in contrast, the r-process stalls at the first or second r-process peak and the heating rate is flatter at late times due to the presence of long-lived radioisotopes. Thus, the Spitzer data provides conditional evidence that GW170817 produced 3rd peak r-process elements. 

Another simple check to this inference is to compare the bolometric light curve to the electron heating rates calculated based on the solar abundance pattern (Figure~\ref{fig:fracheating}).  
The Spitzer detections cannot be explained only by radioactive decay of elements in the first abundance peak as none of them have half-life between  between 10--100\,days. Abundant elements with relevant half-life include $^{89}$Sr, $^{125}$Sn, $^{131}$I, $^{140}$La, $^{141}$Ce, $^{143}$Pr, $^{144}$Ce, $^{156}$Eu, $^{188}$Re, $^{188}$W.
Thus, while the early ground-based data can be explained by many different subsets of r-process elements with different relative abundances, the late-time Spitzer data require the presence of the heaviest r-process elements. 

One caveat here is that we cannot rule out an abundance distribution that cuts off at lanthanum (A=140). However, nuclear reaction network calculations show that only a narrow range in Ye near 0.25 produce lanthanides but not third peak elements (e.g., \citealt{Lippuner2015,Rosswog2018,Holmbeck2018}).  Simulations of NS mergers generally show that the Ye of the total ejecta has a broad distribution, and is not narrowly spiked around a single value like Ye = 0.25 (e.g., \citealt{Perego2014, Sekiguchi2016, Siegel2018, Fernandez2019}). 

Finally, we note that alpha decay and spontaneous fission of the heaviest nuclei could further enhance the late-time infrared emission. Specifically, \citealt{Zhu2018} predict that Californium-254 could boost the mid-IR emission to $-13.2$\,mag at +50\,days for a specific model. However, our measured absolute magnitude at 4.5$\mu$m of $-$11.1\,mag AB at +43\,days is fainter than their predicted contribution from spontaneous fission of Californium-254 by 2\,mag and decays more steeply than their prediction. Perhaps, 
a lower Californium-254 abundance and different bolometric corrections at the two Spitzer epochs could resolve this discrepancy.

\begin{table*}
  \caption{Spitzer mid-IR data on GW\,170817}
\begin{tabular}{llllll}
 \hline
UTC (Phase) & Instrument & Filter & Reference & Mag (Vega) & Mag (AB) \\
 \hline
2017-09-29 (+43\,d) & {\it Spitzer}/IRAC  & 4.5$\mu$m & 2018-05-08 & 18.62  & 21.88 $\pm$ 0.04 ($\pm$ 0.05) \\
2017-10-30 (+74\,d) &  {\it Spitzer}/IRAC & 4.5$\mu$m & 2018-05-08 & 20.60 & 23.86 $\pm$ 0.22 ($\pm$ 0.05)  \\
2017-09-29 (+43\,d) & {\it Spitzer}/IRAC  & 3.6$\mu$m & 2018-05-08 &  $>$20.42 (3$\sigma$) & $>$23.21 (3$\sigma$) \\ 
2017-10-30 (+74\,d) &  {\it Spitzer}/IRAC & 3.6$\mu$m & 2018-05-08 & $>$20.26 (3$\sigma$) & $>$23.05 (3$\sigma$) \\
 \hline
\end{tabular}
 \label{tab:data}
\end{table*}

\section{Conclusions}

In summary, the Spitzer 4.5$\mu$m observations of GW170817 are the latest and reddest detections of the kilonova emission: 21.88 $\pm$ 0.04\,mag at +43\,d and 23.86 $\pm$ 0.22\,mag at +74\,d. The inferred luminosity suggests a broad composition of r-process elements including the heaviest elements in the second and third abundance peak.

We conclude that mid-IR observations are a critical diagnostic of r-process nucleosynthesis and directly constrain the relative composition. This information cannot be gleaned from the near-IR bands that are accessible from the ground. Currently, warm {\it Spitzer} is planned to be online during LIGO-Virgo's third observing run in 2019. We hope that {\it Spitzer} remains online for the entire duration of the observing run. Photometry with {\it Spitzer}, in particular, the late-time bolometric luminosity and slope would uniquely constrain the abundance of the heaviest elements. The warm {\it Spitzer} analysis is currently limited only by the uncertain bolometric corrections as only two bands are available. This uncertainty would be alleviated by future measurements of the full spectral energy distribution in the mid-IR.   

The James Webb Space Telescope (JWST) is planned to be launched soon after the LIGO-Virgo interferometers plan to attain full sensitivity. Monitoring the SED evolution from 1-25$\mu$m would rule out a chunk of parameter space in heating rates. Specifically, the JWST/NIRCAM F444W filter would be able to detect a GW170817-like event at +74\,d as far out as 440\,Mpc in less than 10\,ks.  Furthermore, JWST sensitivity is well-matched to obtaining spectra in the nebular phase that would be a direct diagnostic of the nuclear composition. The JWST/NIRSPEC G395M/F290LP instrument could get a R$\approx$1000 spectrum spanning 2.9-5.1$\mu$m for a GW170817-like event at +43\,d as far out as 92\,Mpc. The JWST/MIRI F1000W could image at 10$\mu$m out to 183\,Mpc. 

Our {\it Spitzer} data decline for GW170817 indicates that elements with a half-life around 14\,days could dominate at late-time. Only a handful of the heaviest elements synthesized by the r-process have half-life between 10\,d--30\,d, e.g.  $^{140}$Ba, $^{143}$Pr, $^{147}$Nd, $^{156}$Eu, $^{191}$Os, $^{223}$Ra, $^{225}$Ra, $^{233}$Pa, $^{234}$Th. Given the lower velocities at late-time and lower line-blending, we may even be able to directly read off line identifications and abundances from future JWST spectra.  

\section*{Acknowledgements}

This work was supported by the GROWTH (Global Relay of Observatories Watching Transients Happen) project funded by the National Science Foundation Partnership in International Research Program under NSF PIRE grant number 1545949.  MMK and DK acknowledge stimulating discussions at KITP; this research was supported in part by the National Science Foundation under Grant No. NSF PHY-1748958. DK is supported in part by the U.S. Department of Energy, Office of Science, Office of Nuclear Physics, under contract number DE-AC02-05CH11231 and DE-SC0017616, and by a SciDAC award DE-SC0018297. This research was supported in part by the Gordon and Betty Moore Foundation through Grant GBMF5076. MMK and EOO thank the United States-Israel Binational Science Foundation for BSF 2016227. We thank the anonymous referee for constructive feedback. We thank B. Metzger, J. L. Barnes, D. Siegel, G. Martinez-Pinedo, M. Wu and R. Surman for valuable discussions.





\appendix

\section{Additional Figure and Table}

\begin{table*}
\caption{Aperture Correction Factors \label{tab:apcor}}
\begin{tabular}{llll}
\hline
 Difference Image & Filter & Photometry Method  & Correction Factor \\
 \hline
 Epoch1-Epoch3 & 3.6$\mu$m & Aperture radius 2.5\,pix, background 4pix-12pix & 1.83 \\
 Epoch2-Epoch3 & 3.6$\mu$m & Aperture radius 2.5\,pix, background 4pix-12pix & 1.84 \\
 Epoch1-Epoch2 & 3.6$\mu$m & Aperture radius 2.5\,pix, background 4pix-12pix & 1.84 \\
 Epoch1-Epoch3 & 4.5$\mu$m & Aperture radius 2.5\,pix, background 4pix-12pix & 1.76 \\
 Epoch2-Epoch3 & 4.5$\mu$m & Aperture radius 2.5\,pix, background 4pix-12pix & 1.76 \\
 Epoch1-Epoch2 & 4.5$\mu$m & Aperture radius 2.5\,pix, background 4pix-12pix & 1.80 \\
 Epoch1-Epoch3 & 4.5$\mu$m & Aperture radius 4.0\,pix, background 4pix-12pix & 1.27 \\
 Epoch1-Epoch3 & 4.5$\mu$m & Aperture radius 6.0\,pix, background 6pix-14pix & 1.13 \\
 Epoch1-Epoch3 & 4.5$\mu$m & Aperture radius 10.0\,pix, background 10pix-20pix & 1.06 \\
 Epoch1-Epoch3 & 4.5$\mu$m & PSF-fit Gaussian FWHM 2.8\,pix & 1.62 \\
 Epoch2-Epoch3 & 4.5$\mu$m & PSF-fit Gaussian FWHM 2.8\,pix & 1.62 \\
 Epoch1-Epoch2 & 4.5$\mu$m & PSF-fit Gaussian FWHM 2.8\,pix & 1.67 \\
 Epoch1-Epoch3 & 4.5$\mu$m & PSF-fit Gaussian FWHM 3.5\,pix & 1.28 \\
 Epoch1-Epoch2 & 4.5$\mu$m & PSF-fit Gaussian FWHM 3.7\,pix & 1.23 \\
 \hline
 \end{tabular}
\end{table*}

 \begin{figure*} 
\centering
\includegraphics[width=0.8\textwidth]{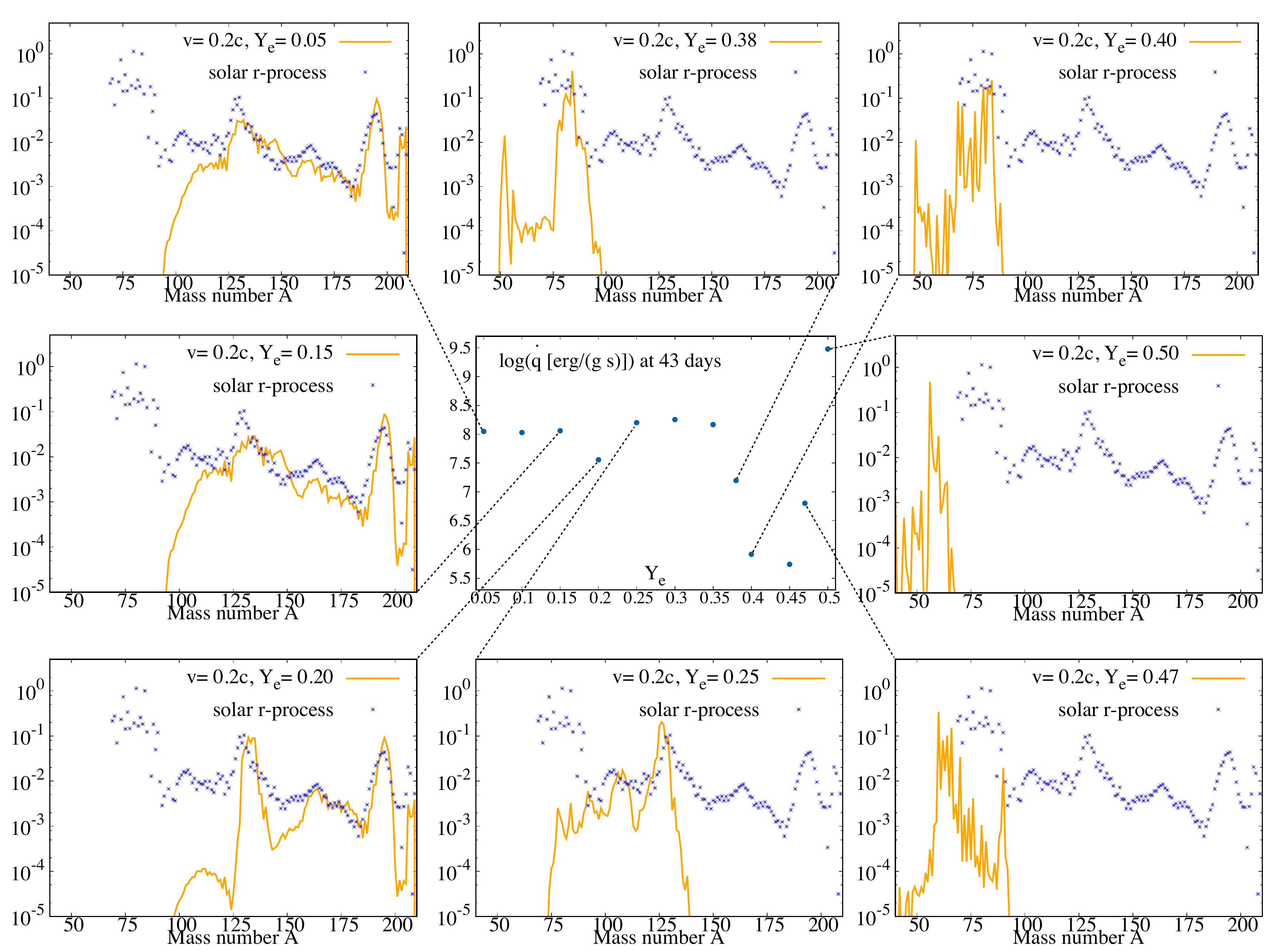}
\caption{Heating rates as a function of composition by running network simulations with fixed entropy (15 k$_{\rm B}$/baryon), fixed velocity (0.2c) and fixed ejecta mass (0.05 M$_{\odot}$). This is the naked energy released before correcting for trapping efficiency and neutrinos \citep{Rosswog2018}.
\label{fig:heating}}
\end{figure*}


\bsp	
\label{lastpage}

\begin{thebibliography}{}
\makeatletter
\relax
\def\mn@urlcharsother{\let\do\@makeother \do\$\do\&\do\#\do\^\do\_\do\%\do\~}
\def\mn@doi{\begingroup\mn@urlcharsother \@ifnextchar [ {\mn@doi@}
  {\mn@doi@[]}}
\def\mn@doi@[#1]#2{\def\@tempa{#1}\ifx\@tempa\@empty \href
  {http://dx.doi.org/#2} {doi:#2}\else \href {http://dx.doi.org/#2} {#1}\fi
  \endgroup}
\def\mn@eprint#1#2{\mn@eprint@#1:#2::\@nil}
\def\mn@eprint@arXiv#1{\href {http://arxiv.org/abs/#1} {{\tt arXiv:#1}}}
\def\mn@eprint@dblp#1{\href {http://dblp.uni-trier.de/rec/bibtex/#1.xml}
  {dblp:#1}}
\def\mn@eprint@#1:#2:#3:#4\@nil{\def\@tempa {#1}\def\@tempb {#2}\def\@tempc
  {#3}\ifx \@tempc \@empty \let \@tempc \@tempb \let \@tempb \@tempa \fi \ifx
  \@tempb \@empty \def\@tempb {arXiv}\fi \@ifundefined
  {mn@eprint@\@tempb}{\@tempb:\@tempc}{\expandafter \expandafter \csname
  mn@eprint@\@tempb\endcsname \expandafter{\@tempc}}}

\bibitem[\protect\citeauthoryear{{Abbott} et~al.,}{{Abbott}
  et~al.}{2017}]{GW170817}
{Abbott} B.~P.,  et~al., 2017, \mn@doi [Physical Review Letters]
  {10.1103/PhysRevLett.119.161101}, \href
  {http://adsabs.harvard.edu/abs/2017PhRvL.119p1101A} {119, 161101}

\bibitem[\protect\citeauthoryear{{Arcavi}}{{Arcavi}}{2018}]{Arcavi2018}
{Arcavi} I.,  2018, \mn@doi [\apjl] {10.3847/2041-8213/aab267}, \href
  {http://adsabs.harvard.edu/abs/2018ApJ...855L..23A} {855, L23}

\bibitem[\protect\citeauthoryear{{Arcavi} et~al.,}{{Arcavi}
  et~al.}{2017}]{Arcavi2017}
{Arcavi} I.,  et~al., 2017, \mn@doi [\nat] {10.1038/nature24291}, \href
  {http://adsabs.harvard.edu/abs/2017Natur.551...64A} {551, 64}

\bibitem[\protect\citeauthoryear{{Barnes} \& {Kasen}}{{Barnes} \&
  {Kasen}}{2013}]{Barnes2013}
{Barnes} J.,  {Kasen} D.,  2013, \mn@doi [\apj] {10.1088/0004-637X/775/1/18},
  \href {http://adsabs.harvard.edu/abs/2013ApJ...775...18B} {775, 18}

\bibitem[\protect\citeauthoryear{{Barnes}, {Kasen}, {Wu}  \&
  {Mart{\'{\i}}nez-Pinedo}}{{Barnes} et~al.}{2016}]{Barnes2016}
{Barnes} J.,  {Kasen} D.,  {Wu} M.-R.,   {Mart{\'{\i}}nez-Pinedo} G.,  2016,
  \mn@doi [\apj] {10.3847/0004-637X/829/2/110}, \href
  {http://adsabs.harvard.edu/abs/2016ApJ...829..110B} {829, 110}

\bibitem[\protect\citeauthoryear{{Coulter} et~al.,}{{Coulter}
  et~al.}{2017}]{Coulter17}
{Coulter} D.~A.,  et~al., 2017, \mn@doi [Science] {10.1126/science.aap9811},
  \href {http://adsabs.harvard.edu/abs/2017Sci...358.1556C} {358, 1556}

\bibitem[\protect\citeauthoryear{{Cowperthwaite} et~al.,}{{Cowperthwaite}
  et~al.}{2017}]{Cowperthwaite17}
{Cowperthwaite} P.~S.,  et~al., 2017, \mn@doi [\apjl]
  {10.3847/2041-8213/aa8fc7}, \href
  {http://adsabs.harvard.edu/abs/2017ApJ...848L..17C} {848, L17}

\bibitem[\protect\citeauthoryear{{Drout} et~al.,}{{Drout}
  et~al.}{2017}]{Drout17}
{Drout} M.~R.,  et~al., 2017, \mn@doi [Science] {10.1126/science.aaq0049},
  \href {http://adsabs.harvard.edu/abs/2017Sci...358.1570D} {358, 1570}

\bibitem[\protect\citeauthoryear{{Evans} et~al.,}{{Evans}
  et~al.}{2017}]{Evans17}
{Evans} P.~A.,  et~al., 2017, \mn@doi [Science] {10.1126/science.aap9580},
  \href {http://adsabs.harvard.edu/abs/2017Sci...358.1565E} {358, 1565}

\bibitem[\protect\citeauthoryear{{Fazio} et~al.,}{{Fazio}
  et~al.}{2004}]{Fazio2004}
{Fazio} G.~G.,  et~al., 2004, \mn@doi [\apjs] {10.1086/422843}, \href
  {http://adsabs.harvard.edu/abs/2004ApJS..154...10F} {154, 10}

\bibitem[\protect\citeauthoryear{{Fern{\'a}ndez}, {Tchekhovskoy}, {Quataert},
  {Foucart}  \& {Kasen}}{{Fern{\'a}ndez} et~al.}{2019}]{Fernandez2019}
{Fern{\'a}ndez} R.,  {Tchekhovskoy} A.,  {Quataert} E.,  {Foucart} F.,
  {Kasen} D.,  2019, \mn@doi [\mnras] {10.1093/mnras/sty2932}, \href
  {http://adsabs.harvard.edu/abs/2019MNRAS.482.3373F} {482, 3373}

\bibitem[\protect\citeauthoryear{{Holmbeck}, {Surman}, {Sprouse}, {Mumpower},
  {Vassh}, {Beers}  \& {Kawano}}{{Holmbeck} et~al.}{2018}]{Holmbeck2018}
{Holmbeck} E.~M.,  {Surman} R.,  {Sprouse} T.~M.,  {Mumpower} M.~R.,  {Vassh}
  N.,  {Beers} T.~C.,   {Kawano} T.,  2018, arXiv e-prints, \href
  {https://ui.adsabs.harvard.edu/\#abs/2018arXiv180706662H} {p.
  arXiv:1807.06662}

\bibitem[\protect\citeauthoryear{{Hotokezaka}, {Sari}  \& {Piran}}{{Hotokezaka}
  et~al.}{2017}]{Hotokezaka17}
{Hotokezaka} K.,  {Sari} R.,   {Piran} T.,  2017, \mn@doi [\mnras]
  {10.1093/mnras/stx411}, \href
  {http://adsabs.harvard.edu/abs/2017MNRAS.468...91H} {468, 91}

\bibitem[\protect\citeauthoryear{{Kasen} \& {Barnes}}{{Kasen} \&
  {Barnes}}{2018}]{Kasen18}
{Kasen} D.,  {Barnes} J.,  2018, preprint, \href
  {http://adsabs.harvard.edu/abs/2018arXiv180703319K} {} (\mn@eprint {arXiv}
  {1807.03319})

\bibitem[\protect\citeauthoryear{{Kasen}, {Badnell}  \& {Barnes}}{{Kasen}
  et~al.}{2013}]{Kasen2013}
{Kasen} D.,  {Badnell} N.~R.,   {Barnes} J.,  2013, \mn@doi [\apj]
  {10.1088/0004-637X/774/1/25}, \href
  {http://adsabs.harvard.edu/abs/2013ApJ...774...25K} {774, 25}

\bibitem[\protect\citeauthoryear{{Kasen}, {Metzger}, {Barnes}, {Quataert}  \&
  {Ramirez-Ruiz}}{{Kasen} et~al.}{2017}]{Kasen17}
{Kasen} D.,  {Metzger} B.,  {Barnes} J.,  {Quataert} E.,   {Ramirez-Ruiz} E.,
  2017, \mn@doi [\nat] {10.1038/nature24453}, \href
  {http://adsabs.harvard.edu/abs/2017Natur.551...80K} {551, 80}

\bibitem[\protect\citeauthoryear{{Kasliwal} et~al.,}{{Kasliwal}
  et~al.}{2017}]{Kasliwal17c}
{Kasliwal} M.~M.,  et~al., 2017, \mn@doi [Science] {10.1126/science.aap9455},
  \href {http://adsabs.harvard.edu/abs/2017Sci...358.1559K} {358, 1559}

\bibitem[\protect\citeauthoryear{{Kulkarni}}{{Kulkarni}}{2005}]{Kulkarni2005}
{Kulkarni} S.~R.,  2005, ArXiv Astrophysics e-prints, \href
  {http://adsabs.harvard.edu/abs/2005astro.ph.10256K} {}

\bibitem[\protect\citeauthoryear{{Lattimer} \& {Schramm}}{{Lattimer} \&
  {Schramm}}{1974}]{Lattimer1974}
{Lattimer} J.~M.,  {Schramm} D.~N.,  1974, \mn@doi [\apjl] {10.1086/181612},
  \href {http://adsabs.harvard.edu/abs/1974ApJ...192L.145L} {192, L145}

\bibitem[\protect\citeauthoryear{{Li} \& {Paczy{\'n}ski}}{{Li} \&
  {Paczy{\'n}ski}}{1998}]{Li1998}
{Li} L.,  {Paczy{\'n}ski} B.,  1998, \mn@doi [\apjl] {10.1086/311680}, \href
  {http://adsabs.harvard.edu/abs/1998ApJ...507L..59L} {507, L59}

\bibitem[\protect\citeauthoryear{{Lippuner} \& {Roberts}}{{Lippuner} \&
  {Roberts}}{2015}]{Lippuner2015}
{Lippuner} J.,  {Roberts} L.~F.,  2015, \mn@doi [\apj]
  {10.1088/0004-637X/815/2/82}, \href
  {http://adsabs.harvard.edu/abs/2015ApJ...815...82L} {815, 82}

\bibitem[\protect\citeauthoryear{{Metzger} et~al.,}{{Metzger}
  et~al.}{2010}]{Metzger2010}
{Metzger} B.~D.,  et~al., 2010, \mn@doi [\mnras]
  {10.1111/j.1365-2966.2010.16864.x}, \href
  {http://adsabs.harvard.edu/abs/2010MNRAS.406.2650M} {406, 2650}

\bibitem[\protect\citeauthoryear{{Mooley} et~al.,}{{Mooley}
  et~al.}{2018}]{Mooley18}
{Mooley} K.~P.,  et~al., 2018, \mn@doi [\nat] {10.1038/nature25452}, \href
  {http://adsabs.harvard.edu/abs/2018Natur.554..207M} {554, 207}

\bibitem[\protect\citeauthoryear{{Peng}, {Ho}, {Impey}  \& {Rix}}{{Peng}
  et~al.}{2010}]{Peng2010}
{Peng} C.~Y.,  {Ho} L.~C.,  {Impey} C.~D.,   {Rix} H.-W.,  2010, \mn@doi [\aj]
  {10.1088/0004-6256/139/6/2097}, \href
  {http://adsabs.harvard.edu/abs/2010AJ....139.2097P} {139, 2097}

\bibitem[\protect\citeauthoryear{{Perego}, {Rosswog}, {Cabez{\'o}n},
  {Korobkin}, {K{\"a}ppeli}, {Arcones}  \& {Liebend{\"o}rfer}}{{Perego}
  et~al.}{2014}]{Perego2014}
{Perego} A.,  {Rosswog} S.,  {Cabez{\'o}n} R.~M.,  {Korobkin} O.,
  {K{\"a}ppeli} R.,  {Arcones} A.,   {Liebend{\"o}rfer} M.,  2014, \mn@doi
  [\mnras] {10.1093/mnras/stu1352}, \href
  {http://adsabs.harvard.edu/abs/2014MNRAS.443.3134P} {443, 3134}

\bibitem[\protect\citeauthoryear{{Perley} et~al.,}{{Perley}
  et~al.}{2016}]{Perley2016}
{Perley} D.~A.,  et~al., 2016, \mn@doi [\apj] {10.3847/0004-637X/817/1/8},
  \href {http://adsabs.harvard.edu/abs/2016ApJ...817....8P} {817, 8}

\bibitem[\protect\citeauthoryear{{Pian} et~al.,}{{Pian}
  et~al.}{2017}]{Pian2017}
{Pian} E.,  et~al., 2017, \mn@doi [\nat] {10.1038/nature24298}, \href
  {http://adsabs.harvard.edu/abs/2017Natur.551...67P} {551, 67}

\bibitem[\protect\citeauthoryear{{Rosswog}, {Feindt}, {Korobkin}, {Wu},
  {Sollerman}, {Goobar}  \& {Martinez-Pinedo}}{{Rosswog}
  et~al.}{2017}]{Rosswog2017}
{Rosswog} S.,  {Feindt} U.,  {Korobkin} O.,  {Wu} M.-R.,  {Sollerman} J.,
  {Goobar} A.,   {Martinez-Pinedo} G.,  2017, \mn@doi [Classical and Quantum
  Gravity] {10.1088/1361-6382/aa68a9}, \href
  {http://adsabs.harvard.edu/abs/2017CQGra..34j4001R} {34, 104001}

\bibitem[\protect\citeauthoryear{{Rosswog}, {Sollerman}, {Feindt}, {Goobar},
  {Korobkin}, {Wollaeger}, {Fremling}  \& {Kasliwal}}{{Rosswog}
  et~al.}{2018}]{Rosswog2018}
{Rosswog} S.,  {Sollerman} J.,  {Feindt} U.,  {Goobar} A.,  {Korobkin} O.,
  {Wollaeger} R.,  {Fremling} C.,   {Kasliwal} M.~M.,  2018, \mn@doi [\aap]
  {10.1051/0004-6361/201732117}, \href
  {http://adsabs.harvard.edu/abs/2018A%26A...615A.132R} {615, A132}

\bibitem[\protect\citeauthoryear{{Sekiguchi}, {Kiuchi}, {Kyutoku}, {Shibata}
  \& {Taniguchi}}{{Sekiguchi} et~al.}{2016}]{Sekiguchi2016}
{Sekiguchi} Y.,  {Kiuchi} K.,  {Kyutoku} K.,  {Shibata} M.,   {Taniguchi} K.,
  2016, \mn@doi [\prd] {10.1103/PhysRevD.93.124046}, \href
  {http://adsabs.harvard.edu/abs/2016PhRvD..93l4046S} {93, 124046}

\bibitem[\protect\citeauthoryear{{Siegel} \& {Metzger}}{{Siegel} \&
  {Metzger}}{2018}]{Siegel2018}
{Siegel} D.~M.,  {Metzger} B.~D.,  2018, \mn@doi [\apj]
  {10.3847/1538-4357/aabaec}, \href
  {http://adsabs.harvard.edu/abs/2018ApJ...858...52S} {858, 52}

\bibitem[\protect\citeauthoryear{{Skrutskie} et~al.,}{{Skrutskie}
  et~al.}{2006}]{2MASS}
{Skrutskie} M.~F.,  et~al., 2006, \mn@doi [\aj] {10.1086/498708}, \href
  {http://adsabs.harvard.edu/abs/2006AJ....131.1163S} {131, 1163}

\bibitem[\protect\citeauthoryear{{Smartt} et~al.,}{{Smartt}
  et~al.}{2017}]{Smartt17}
{Smartt} S.~J.,  et~al., 2017, \mn@doi [\nat] {10.1038/nature24303}, \href
  {http://adsabs.harvard.edu/abs/2017Natur.551...75S} {551, 75}

\bibitem[\protect\citeauthoryear{{Soares-Santos} et~al.}{{Soares-Santos}
  et~al.}{2017}]{SoaresSantos2017}
{Soares-Santos} M.,  et~al., 2017, \mn@doi [\apjl] {10.3847/2041-8213/aa9059},
  848, in press

\bibitem[\protect\citeauthoryear{{Tanaka} \& {Hotokezaka}}{{Tanaka} \&
  {Hotokezaka}}{2013}]{Tanaka2013}
{Tanaka} M.,  {Hotokezaka} K.,  2013, \mn@doi [\apj]
  {10.1088/0004-637X/775/2/113}, \href
  {http://adsabs.harvard.edu/abs/2013ApJ...775..113T} {775, 113}

\bibitem[\protect\citeauthoryear{{Villar} et~al.,}{{Villar}
  et~al.}{2018}]{Villar18}
{Villar} V.~A.,  et~al., 2018, preprint, \href
  {http://adsabs.harvard.edu/abs/2018arXiv180508192V} {} (\mn@eprint {arXiv}
  {1805.08192})

\bibitem[\protect\citeauthoryear{{Waxman}, {Ofek}, {Kushnir}  \&
  {Gal-Yam}}{{Waxman} et~al.}{2017}]{Waxman17}
{Waxman} E.,  {Ofek} E.,  {Kushnir} D.,   {Gal-Yam} A.,  2017, preprint, \href
  {http://adsabs.harvard.edu/abs/2017arXiv171109638W} {} (\mn@eprint {arXiv}
  {1711.09638})

\bibitem[\protect\citeauthoryear{{Werner} et~al.,}{{Werner}
  et~al.}{2004}]{Werner2004}
{Werner} M.~W.,  et~al., 2004, \mn@doi [\apjs] {10.1086/422992}, \href
  {http://adsabs.harvard.edu/abs/2004ApJS..154....1W} {154, 1}

\bibitem[\protect\citeauthoryear{{Wollaeger} et~al.,}{{Wollaeger}
  et~al.}{2018}]{Wollaeger2017}
{Wollaeger} R.~T.,  et~al., 2018, \mn@doi [\mnras] {10.1093/mnras/sty1018},
  \href {http://adsabs.harvard.edu/abs/2018MNRAS.478.3298W} {478, 3298}

\bibitem[\protect\citeauthoryear{{Wu}, {Barnes}, {Martinez-Pinedo}  \&
  {Metzger}}{{Wu} et~al.}{2018}]{Wu2018}
{Wu} M.-R.,  {Barnes} J.,  {Martinez-Pinedo} G.,   {Metzger} B.~D.,  2018,
  preprint, \href {http://adsabs.harvard.edu/abs/2018arXiv180810459W} {}
  (\mn@eprint {arXiv} {1808.10459})

\bibitem[\protect\citeauthoryear{{Zackay}, {Ofek}  \& {Gal-Yam}}{{Zackay}
  et~al.}{2016}]{zogy}
{Zackay} B.,  {Ofek} E.~O.,   {Gal-Yam} A.,  2016, \mn@doi [\apj]
  {10.3847/0004-637X/830/1/27}, \href
  {http://adsabs.harvard.edu/abs/2016ApJ...830...27Z} {830, 27}

\bibitem[\protect\citeauthoryear{{Zhu} et~al.,}{{Zhu} et~al.}{2018}]{Zhu2018}
{Zhu} Y.,  et~al., 2018, preprint, \href
  {http://adsabs.harvard.edu/abs/2018arXiv180609724Z} {} (\mn@eprint {arXiv}
  {1806.09724})

\makeatother
\end{thebibliography}
\end{document}